\documentclass[sigconf]{acmart}

\usepackage{booktabs} 
\usepackage{enumitem}
\usepackage{balance}

\copyrightyear{2017} 
\acmYear{2017} 
\setcopyright{acmcopyright}
\acmConference{DLRS 2017}{August 27, 2017}{Como, Italy}\acmPrice{15.00}\acmDOI{10.1145/3125486.3125495}
\acmISBN{978-1-4503-5353-3/17/08}

\begin{document}
\title{Comparing Neural and Attractiveness-based Visual Features for Artwork Recommendation}

\author{Vicente Dominguez}
\affiliation{%
  \institution{Pontificia Universidad Catolica de Chile}
  \city{Santiago} 
  \state{Chile} 
}
\email{vidominguez@uc.cl}

\author{Pablo Messina}
\affiliation{%
  \institution{Pontificia Universidad Catolica de Chile}
  \city{Santiago} 
  \state{Chile} 
}
\email{pamessina@uc.cl}

\author{Denis Parra}
\affiliation{%
  \institution{Pontificia Universidad Catolica de Chile}
  \city{Santiago} 
  \state{Chile} 
}
\email{dparra@ing.puc.cl}

\author{Domingo Mery}
\affiliation{%
  \institution{Pontificia Universidad Catolica de Chile}
  \city{Santiago} 
  \state{Chile} 
}
\email{dmery@ing.puc.cl}

\author{Christoph Trattner}
\affiliation{%
  \institution{MODUL University Vienna}
  \city{Vienna} 
  \state{Austria} 
}
\email{christoph.trattner@modul.ac.at}

\author{Alvaro Soto}
\affiliation{%
  \institution{Pontificia Universidad Catolica de Chile}
  \city{Santiago} 
  \state{Chile} 
}
\email{asoto@ing.puc.cl}

\renewcommand{\shortauthors}{Dominguez et al.}

\begin{abstract}
Advances in image processing and computer vision in the latest years have brought about the use of visual features in artwork recommendation. Recent works have shown that visual features obtained from pre-trained deep neural networks (DNNs) perform extremely well for recommending digital art. Other recent works have shown that explicit visual features (EVF) based on attractiveness can perform well in preference prediction tasks, but no previous work has compared DNN features versus specific attractiveness-based visual features (e.g. brightness, texture) in terms of recommendation performance. In this work, we study and compare the performance of DNN and EVF features for the purpose of physical artwork recommendation using transactional data from UGallery, an online store of physical paintings. In addition, we perform an exploratory analysis to understand if DNN embedded features have some relation with certain EVF. Our results show that DNN features outperform EVF, that certain EVF features are more suited for physical artwork recommendation and, finally, we show evidence that certain neurons in the DNN might be partially encoding visual features such as brightness, providing an opportunity for explaining recommendations based on visual neural models.

\end{abstract}

%


\maketitle

\section{Introduction}

In the latest five years, the area of computer vision has been revolutionized by deep neural networks (DNN), where the use of visual neural embeddings 
from pre-trained convolutional neural networks has increased by orders of magnitude the performance on tasks such as image classification \cite{krizhevsky2012imagenet} or scene identification \cite{sharif2014cnn}.
In the area of recommender systems, a few works have exploited neural visual embeddings for recommendation, such as \cite{mcauley2015image, he2016vista,he2016vbpr,1704.06109}. Among these works, we are particularly interested in the use of visual features for recommending art \cite{he2016vista}.
The online artwork market is booming due to the influence of social media and new consumption behaviors of millennials \cite{forbes2}, but the main works for recommending art date  more than 8 years \cite{aroyo2007personalized} and they did not utilize visual features for recommendation.

In a recent work, He et al. \cite{he2016vista} introduced a recommender method for digital art employing ratings, social features and pre-trained visual DNN embeddings with very good results. However, they did not use explicit visual features (EVF) such as brightness, color or texture. Using only latent embeddings (from DNNs) affects model transparency and explainability of recommendations, which can in turn hinder the user acceptance of the suggestions \cite{verbert2013visualizing,konstan2012recommender}. In addition, their work focused on digital art, and in the present work we are interested in recommending physical artworks (paintings).
In a more recent work \cite{messina2017}, we compared the performance of visual DNN features versus art metadata and explicit visual features (EVF: colorfulness, brightness, etc.) for recommendation of physical artworks. We showed that visual features (DNN and EVF) outperform manually-curated metadata. However, we did not analyze which specific visual features are more important in the artwork recommendation task. Furthermore, although previous research have studied what is being encoded by neurons in a DNN \cite{nguyen2016multifaceted,zeiler2014visualizing}, to the best of our knowledge, no previous work has investigated a link between latent DNN visual features and attractiveness visual features such as those investigated by San Pedro et al. \cite{sanpedro2009}  (colorfulness, brightness, etc.). Understanding what visual neural models are encoding could help in transparency and explainability of recommender systems \cite{verbert2013visualizing}.

%
%


\paragraph{Contribution} In this work we compare the performance of 8 explicit visual features (EVF) for the task of recommending physical paintings from an online store, UGallery\footnote{http://www.UGallery.com}. Our results indicate that the combination of these features offers the best performance, but it also highlights that features like entropy and contrast contribute more than features like colorfulness to the recommendation task. Moreover, an exploratory analysis provides evidence that certain latent features from the DNN visual embedding might be partially related to explicit visual features. This result could be eventually utilized in explaining recommendations with black-box neural visual models.

\section{Related Work}

Here we survey some works using visual features obtained from pre-trained deep neural networks for recommendation tasks. McAuley et al. \cite{mcauley2015image} introduced an image-based recommendation system based on styles and substitutes for clothing using visual embeddings pre-trained on a large-scale dataset obtained from Amazon.com. Recently, He et al. \cite{he2016vbpr} went further in this line of research and introduced a visually-aware matrix factorization approach that incorporates
visual signals (from a pre-trained DNN) into predictors of people's opinions.
The latest work by He et al. \cite{he2016vista} deals  with visually-aware artistic recommendation, building a model which combines ratings, social signals and visual features.
Deldjoo et al. \cite{1704.06109} compared visual DNN and explicit (stylistic) visual features for movie recommendation, and they found the explicit visual features more informative than DNN features for their recommendation task.

Unlike these previous works, we compare in this article neural and specific explicit visual features (such as brightness or contrast) for physical painting recommendation, and we also explore the potential relation between these two types of features (DNN vs. EVF features).

\section{Problem Description \& Dataset}
\label{sec:recproblem}

The online web store \textit{UGallery} supports emergent artists by helping them to sell their original paintings online. In this work, we study content-based top-n recommendation based on sets of visual features extracted directly from images. In order to perform personalized recommendations, we create a user profile based on paintings already bought by a user, and based on this model we attempt to predict the next paintings the user will buy. In this work, we focus on comparing different sets of visual features, in order to understand which ones could provide a better recommendation.


\textbf{Dataset}. UGallery shared with us an anonymized dataset of $1,371$ users, $3,490$ paintings (with their respective images) and $2,846$ purchases (transactions) of paintings, where all users have made at least one transaction. In average, each user has bought 2-3 items in the latest years\footnote{Our collaborators at UGallery requested us not to disclose the exact dates where the data was collected.}.

\begin{figure}[!ht]
\centering
    	\scalebox{1.0}{
        \includegraphics[width=\linewidth]{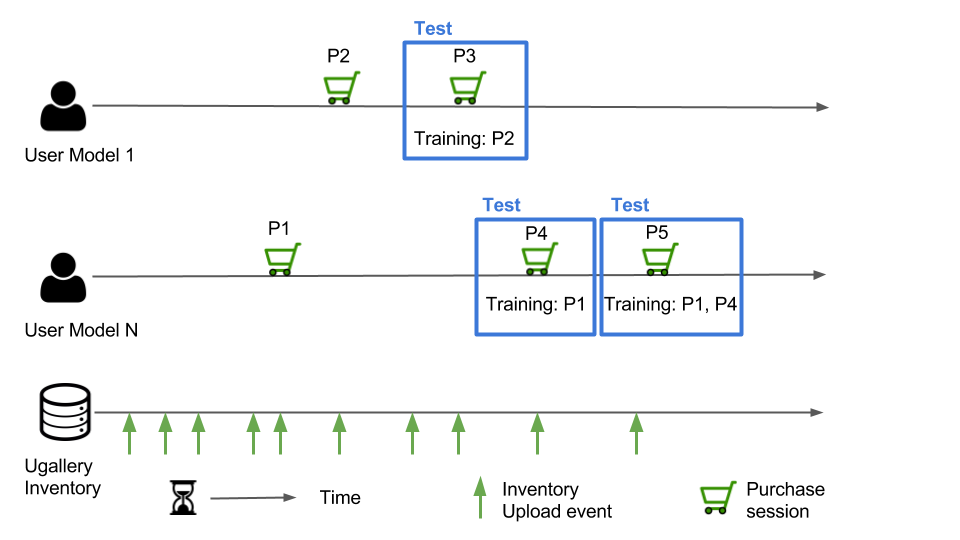}
        }
    \vspace*{-1em}
        \caption{Evaluation procedure. Each surrounding box represents a test, where we predict the items of the purchase session. In the figure, we predict which artworks bought User 1 in purchase P3. `Training:P2' means that we used items from purchase session P2 to train the model.}
\label{fig:evaluation}
\vspace*{-3mm}
\end{figure}

\section{Features \& Recommendation Method}

Since we are using a content-based approach to produce recommendations, we first describe the visual features extracted from images, and then we describe how we used them to make recommendations.

\textbf{Visual Features}. For each image representing a painting in the dataset we obtain features from a pre-trained AlexNet DNN \cite{krizhevsky2012imagenet}, which outputs a vector of $4,096$ dimensions, the fc6 layer. This network was trained with Caffe~\cite{jia2014caffe} using the ImageNet ILSVRC 2012 dataset~\cite{krizhevsky2012imagenet}. We also tested a pre-trained VGG16 \cite{simonyan2014very} model, but the results were not significantly different so we do not report them in this article.

We also obtain a vector of explicit visual features of attractiveness, using the OpenCV software library\footnote{http://opencv.org/}, based on the work of San Pedro et al. \cite{sanpedro2009}: brightness, saturation, sharpness, entropy, RGB-contrast, colorfulness and naturalness. A more detailed description of these features:

\begin{itemize}[leftmargin=*]
\item \textit{Brightness}: It measures the level of luminance of an image. For images in the \textit{YUV} color space, we obtain the average of the luminance component \textit{Y}.

\item \textit{Saturation}: It measures the vividness of a picture. For images in the \textit{HSV} or \textit{HSL} color space, we obtain the average of the saturation component \textit{S}.

\item \textit{Sharpness}: It measures how detailed is the image.

\item \textit{Colorfulness}: It measures how distant are the colors from the gray color.

\item \textit{Naturalness}: It measures how natural is the picture, grouping the pixels in Sky, Grass and Skins pixels and applying the formula in \cite{sanpedro2009}.

\item  \textit{RGB-contrast}: Measures the variance of luminance in
the RGB color space.

\item \textit{Entropy}: Shannon's entropy is calculated, applied to the histogram of values of every pixel in grayscale used as a vector. The histogram is used as the distribution to calculate the entropy.

\item \textit{Local Binary Patterns}: Although this is not actually an ``explicit'' visual feature, it is a traditional baseline in several computer vision tasks \cite{ojala1996comparative}, so we test it for recommendations too.

\end{itemize}

%
%
\begin{figure*}[h!]
    \centering
    \includegraphics[width=\linewidth]{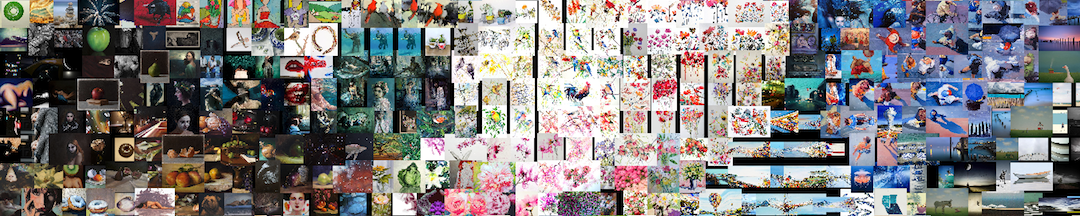}
    \caption{Partial t-SNE map of the Ugallery dataset using the image embeddings from AlexNet DNN. This map helps to show groups of similar images as encoded with the AlexNet embedding. The grid is made using Jonker-Volgenant algorithm \cite{jonker1987shortest}.}
    \label{fig:ugallery-embedding}
\end{figure*}
%
%

\textbf{Recommendation Method}. In order to produce a content-based recommendation list of $k$ artworks for a user $u$, we follow the following procedure: For every item in the current inventory we: (1) calculate its similarity to each item in the user's model, then (2) these single similarities are aggregated into a single score as either the sum or the maximum of all these similarities, and finally (3) the items are sorted by their scores and the top $k$ items in the list are recommended. Formally, given a user $u$ who has consumed a set of artworks $I_u$, and an arbitrary artwork $i$ from the inventory, the score of this item $i$ to be recommended to $u$ is:

\begin{equation}
score(i,u) = \sum_{j \in I_u} sim(V_i,V_j)
\label{eq:DNN}
\end{equation}

where $V_z$ is the feature vector embedding of the painting $z$ obtained with the Alexnet or from the EVFs. Moreover, the similarity function used was cosine similarity, expressed as:

\begin{equation}
sim(V_i,V_j) = cos(V_i,V_j) = \frac{V_i \cdot V_j}{ \lVert{ V_i} \rVert{} \lVert{V_j}\rVert{}}
\label{eq:cosine1}
\end{equation}

\section{Evaluation Method}

Our protocol is based on Macedo {\it et al.} \cite{macedo2015context} to evaluate a recommender system in a time-based manner, and it is presented in Figure \ref{fig:evaluation}. We attempt to predict the items purchased in every transaction, where the training set contains all the artworks previously bought by a user just before making the transaction to be predicted. Users who have purchased exactly one artwork were considered as \emph{cold start} users, and we remove them for this evaluation. After this filtering, our datasets ends up with 365 people who bought more than a single item, who performed a total of 1,629 transactions (i.e. we conducted 1,629 evaluation tests with each algorithm).

\paragraph{Metrics}
As suggested by Cremonesi et al. \cite{Cremonesi2010} for Top-N recommendation, we used $recall@k$ and $precision@k$, as well as nDCG, a commonly used metric for ranking evaluation \cite{manning2008introduction}.


\setlength{\tabcolsep}{2pt}
\begin{table}[t]
\centering
\caption{Results of the recommendation task using a diverse set of latent (DNN) and explicit visual features (EVF).}
\label{tab:results}
\scalebox{0.85}
{
	\begin{tabular}{@{\extracolsep{0pt}}lrrrrrr}
		\hline
		 name               &   ndcg@5 &   ndcg@10 &   rec@5 &   rec@10 &   prec@5 &   prec@10 \\
		\hline
		 DNN                &   \textbf{.0884} &    \textbf{.1027} &  \textbf{.1180} &   \textbf{.1598} &   \textbf{.0290} &    \textbf{.0204} \\
		 EVF (all features) &   .0344 &    .0459 &  .0547 &   .0885 &   .0127 &    .0111 \\
         EVF (all, except LBP)  &   .0370 &    .0453 &  .0585 &   .0826 &   .0152 &    .0109 \\ \hline
    	 EVF (LBP)          &   .0292 &    .0388 &  .0431 &   .0715 &   .0103 &    .0085 \\
		 EVF (brightness)   &   .0048 &    .0083 &  .0080 &   .0186 &   .0035 &    .0031 \\
		 EVF (colorfulness) &   .0020 &    .0052 &  .0033 &   .0126 &   .0008 &    .0017 \\
		 EVF (contrast)     &   .0079 &    .0102 &  .0149 &   .0220 &   .0033 &    .0025 \\
		 EVF (entropy)      &   .0088 &    .0098 &  .0110 &   .0137 &   .0026 &    .0019 \\
		 EVF (naturalness)  &   .0062 &    .0110 &  .0108 &   .0248 &   .0026 &    .0032 \\
		 EVF (saturation)   &   .0055 &    .0084 &  .0096 &   .0187 &   .0021 &    .0020 \\
		 EVF (sharpness)    &   .0063 &    .0085 &  .0117 &   .0178 &   .0024 &    .0021 \\
		\hline
	\end{tabular}

	\begin{tabular}{lrrrrrr}

	\end{tabular}

}
\end{table}

\section{Results}

Table \ref{tab:results} presents the results, which we summarize in three points:
\begin{itemize}[leftmargin=*]
\item  AlexNet DNN features perform better than those based on EVF, either combined or isolated. Figure \ref{fig:ugallery-embedding} presents a sample of the t-SNE map \cite{maaten2008visualizing} made from the AlexNet DNN visual features, which shows well-defined clusters of images.  This result reflects the current state-of-the-art of deep neural networks in computer vision, but as already stated, the lack of transparency of DNN visual features can hinder the user acceptance of these recommendations due to difficulties to explain these recommendations \cite{konstan2012recommender,verbert2013visualizing}.

\item Combining EVF features improves their performance compared to using them isolated, but in some cases it is detrimental. For instance, the combination {\it EVF (all, except LBP)} yields better ndcg@5, recall@5 and precision@5 than {\it EVF (all features)}.

\item By comparing isolated EVF features, {\em LBP} performs the best because it encodes texture patterns and local
contrast very well, however its explanation is more complex than image {\em brightness} or {\em contrast}.
Considering only the 7 original features proposed by SanPedro et al. \cite{sanpedro2009} we observe that {\em entropy}, {\em contrast}, and {\em naturalness} perform consistently well, as well as {\em brightness} in terms of precision. On the other side, {\em colorfulness} does not seem to have a significant impact on providing good recommendations.
\end{itemize}

\subsection{Relation between DNN and EVF features}
We also explored, by a correlation analysis, the potential relation between each of the $4,096$ AlexNet features and the isolated EVFs, results are shown in Table \ref{tab:correlations}. We found that brightness has a significant positive correlation with AlexNet feature $F[598]$ ($\rho=.58$) as well as a significant negative correlation with AlexNet feature $F[3916]$ ($\rho=-.5$). The opposite is found with the feature naturalness, which largest positive and negative correlations are ($rho_{max}=.008$) and ($rho_{min}=-.002$), meaning that no single AlexNet DNN neuron seems to be explicitly encoding naturalness.

%
%
\begin{figure*}[t!]
    \centering
    \includegraphics[width=\linewidth]{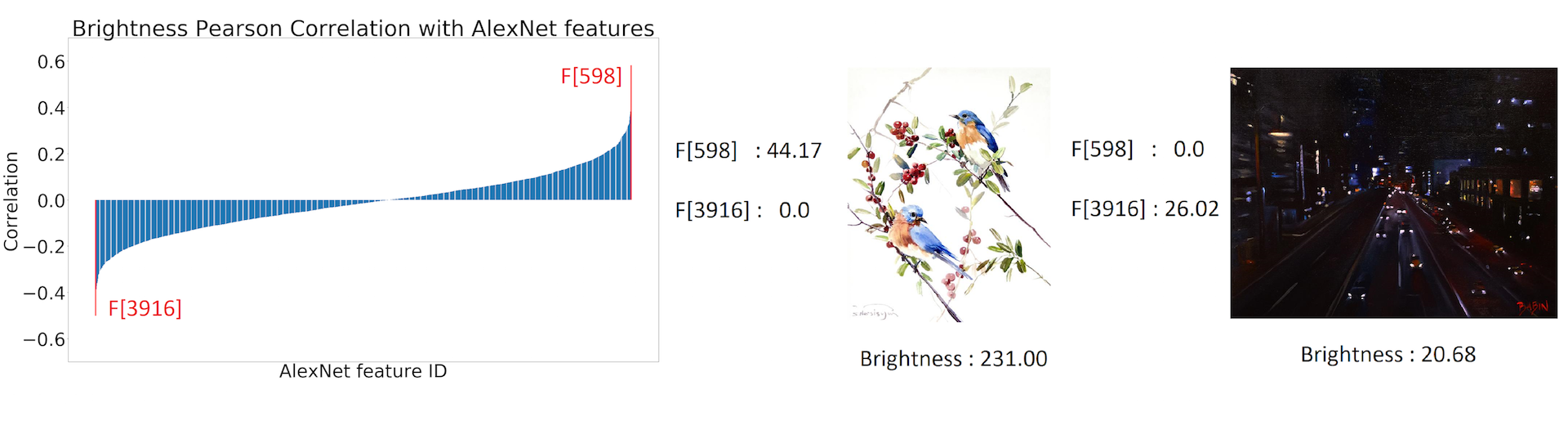}
    \caption{The plot at the left shows the correlation of each AlexNet feature with the visual feature \texttt{Brightness}, highlighting features with most negative ($F[3916]=-.5$) and positive correlations ($F[598]=.58$). Example images at the right side show evidence of this correlation. Brightness values range between $[0..250]$, $F[598]$ between $[0..49.66]$, and $F[3916]$ between $[0..38.79]$. }
    \label{fig:correlation}
\end{figure*}
%
%

Figure \ref{fig:correlation} shows a plot with correlations of {\em brigthness} to all AlexNet dimensions, sorted from the smallest $F[3916]$ to the largest $F[598]$. This figure also presents sampled images that support this analysis, where two images with high and low brightness, also show respective high and low values in the AlexNet features $F[598]$ and the opposite for $F[3916]$.

{\it Limitations}. It is important to note that this is only an exploratory analysis and generalizability should be considered carefully. Nevertheless,  the high correlation of {\em brightness} and the small correlation of {\em naturalness} provides a hint towards what types of features are not being explicitly encoded in the single neurons of the fc6 layer in the AlexNet DNN.

\setlength{\tabcolsep}{2pt}
\renewcommand{\arraystretch}{0.95}
\begin{table}[t!]
\centering
\caption{Maximum and minimum correlation between attractiveness-based visual features and the 4,096 AlexNet embeddings, indicating the respective dimension index.}
\label{tab:correlations}
\scalebox{1}{
\begin{tabular}{lcccc}
\toprule
Visual feature      & $max(corr.)$ & $index(F_{max}) $ & $min(corr.)$  & $index(F_{min})$ \\ \midrule
brightness   & {\bf 0.5815}           & 598         & {\bf -0.5020}          & 3916        \\
sharpness    & 0.4682           & 4095        & -0.3668          & 1025        \\
saturation   & 0.3645           & 445         & -0.5019          & 2370        \\
colorfulness & 0.3034           & 1014        & -0.3383          & 2410        \\
entropy      & 0.3410           & 286         & -0.4369          & 3499        \\
contrast     & 0.3941           & 469         & -0.2799          & 3761        \\
naturalness  & 0.0852           & 2626        & -0.0256          & 2499       \\ \bottomrule
\end{tabular}
}
\end{table}

\section{Conclusion and Future Work}
In this article we have investigated the impact of latent (DNN) and explicit (EVF) visual features on artwork recommendation. Our results support that DNN features outperform explicit visual features. Although one previous work on movie recommendation found the opposite \cite{1704.06109} --i.e. EVF being more informative than DNN visual features--, we argue that the domain differences (paintings vs. movies) and the fact that we use images rather than video might explain the difference, but further research is needed in this aspect.

We also show that some EVF contribute more information for the artwork recommendation task, such as {\em entropy}, {\em contrast}, {\em naturalness} and {\em brightness}. On the other side, {\em colorfulness} is the less informative visual feature for this task.

Moreover, by a correlation analysis we showed that {\em brightness}, {\em saturation} and {\em entropy} are significantly correlated with some DNN embedding features, while naturalness is poorly correlated. This preliminary results should be further studied with the aim of improving the explainability of these black-box models \cite{verbert2013visualizing}.

In future work, we will study the use of other visual embeddings which have shown good results in computer vision tasks, such as GoogleNet \cite{szegedy2015going}. In addition, we will conduct a user study to investigate ways to explain image recommendations using the relations found between visual DNN features and EVF.

\begin{acks}
 We thank Alex Farkas and Stephen Tanenbaum from UGallery for sharing the dataset and answering our questions. We also thank Felipe Cortes, PUC Chile student who implemented a web tool to visualize the DNN embedding. Authors Vicente Dominguez, Pablo Messina and Denis Parra are
 supported by the Chilean research agency \grantsponsor{00}{Conicyt}{}, under Grant {\em Fondecyt Iniciacion}
 No.:~\grantnum{00}{11150783}.

\end{acks}

\balance
\bibliographystyle{ACM-Reference-Format}
\bibliography{sigproc} 

\end{document}